\documentstyle[epsfig]{l-aa}

\begin{document}

   \thesaurus{10         
              (03.13.2;  
               05.01.1;  
               05.05.2;  
               07.03.1;  
               07.19.2)} 
   \title{Kuiper Belt searches from the Palomar 5-m telescope}

   \author{Brett Gladman\inst{1}
	 	\and
	   JJ Kavelaars\inst{2}
          }

   \institute{Dept. of Astronomy
\thanks{
Visiting Astronomers: Mount Palomar (BG and JJK) and Canada-France-Hawaii 
Telescope (JJK).
Observations at the Palomar Observatory were made as part of a continuing 
collaborative agreement between the California Institue of Technology
and Cornell University.
The Canada-France-Hawaii Telescope is operated by
the National Research Council of Canada, le Centre National de
la Recherche Scientifique de France, and the University of Hawaii. },
              406 Space Sciences Bldg.,
              Cornell University,
              Ithaca, NY, 14853, USA
	\and
	      Dept. of Physics and Astronomy$^*$,
	      Queen's University,
	      Kingston, ON,
	      K7L 3N6, CANADA
             }

   \date{Received August ??, 1996; accepted ???? ??, 1996}

   \maketitle

   \begin{abstract}

Motivated by a desire to understand the size distributon of
the Kuiper Belt, an observing program was conducted at the
Mount Palomar 5-m telescope from 1994-1996.
The observations consisted of follow-up observations of
known objects (in order to improve their
very indeterminate orbits), and deep exposures on a single
field to search for small objects below the limiting magnitudes
of other surveys.
Eighteen object recoveries were successfully obtained over the course
of the follow-up program.
Data reduction of the deep fields consisted of a software
recombination of many fields shifted at different angular
rates in order to detect objects at differing heliocentric
distances.
We set an upper limit of $<1$ object per 0.05 square degrees
in the ecliptic brighter than magnitude $R\simeq25$.
The lack of detected objects in this work serves to help constrain 
the number density of Kuiper Belt comets.

   \end{abstract}

%

\section{Introduction and Motivation}

  Observational searches for objects in the outer solar system,
and especially in the trans-Neptunian region, began to
produce consistent results beginning in 1992 with the discovery of 
the first so-called Kuiper Belt object 1992 QB1 (Jewitt and Luu 
1993).
Since then, approximately 3 dozen Kuiper Belt objects (KBOs hereafter) 
have been found (see Stern 1996 for a review), ranging in apparent 
magnitude from about 22 to 24.6 in R-band, and heliocentric distance 
from 30 -- 45 AU.
Assuming a comet-like albedo of $p=0.04$, these objects have 
diameters ranging from 100 to 300 km, although it cannot be
ruled that the brightness variation could be due to differing 
albedos rather than sizes.

The size distribution of KBOs is of great interest.
Although originally it had been hoped that the population might
be collisionless (and thus might hold the signature of the formation 
process), recent work (Stern 1995) has shown that the collisional 
effects cannot be neglected over 4.5 Byr.
However, knowledge of the size distribution is still important for 
understanding the link between the Kuiper Belt and both the
short-period comets (Levison and Duncan 1994) and Pluto 
(Stern 1996).
After this Palomar search program was begun, the HST results
of Cochran et al.~(1995) provided another stong motivation (discussed
below) by statistically detecting a very large population of
Kuiper Belt comets.

The goal of this program was to find small KBOs rather
than more objects with diameters of a few hundred kilometers.
Instead of searching large areas of sky to limiting magnitudes of
$R\simeq 23$ -- 23.5, the intent was to concentrate on a single
field (for each observing run) and integrate for 4--6 hours to
reach a limiting magnitude of $R \simeq 26$.
In essence, the hope was that a power law increase in
the number of objects with decreasing magnitude would dominate
the loss due to searching a single field.

Figure 1 shows a compilation of previous results, adapted from 
a figure from Irwin et al.~(1995), whose assumptions we
adopt below.
The plot shows the number of outer solar system objects as a
function of absolute $H_R$ magnitude.
The $H_R$ magnitude is the apparent magnitude of the object if
it were 1 AU from both the Sun and Earth; we can use
$R = H_R + 10 \log r_{AU} \simeq H_R + 16$ for objects at 40 AU.
The cumulative number $N(< H_R)$ of objects brighter than some
specified $H_R$ magnitude is computed from an estimated projected
sky density by assuming a Kuiper belt of width $\pm 10^\circ$ extending
uniformly around the ecliptic.
Fig.~1 shows the results of direct searches in the Kuiper belt 
(solid symbols), and model-dependent limits (open symbols) based 
on the conversion of Kuiper Belt objects into Centaurs.
The data sets connected by lines on the left half of the figure
are upper limits of various surveys described in Irwin et al.~(1995).
The Jewitt and Luu (JL) survey provides a direct detection of
6 objects per sq.~degree, and model dependent limits on smaller
objects through the non-detection of Centaurs.
The Spacewatch program provides a model-dependent data point due
to their observations of Centaurs; Kowal's discovery of Chiron 
provides another.
The HST observations are best interpreted (Levison, 1996, private
communication) as detecting 2:3 Neptune librators near
their perihelia at $\simeq 33$ AU, which implies $N \sim
2 \times 10^8$ objects brighter than $H_R=12.8$, as is plotted
on Fig.~1.

%
   \begin{figure}[htbp]
\epsfig{file=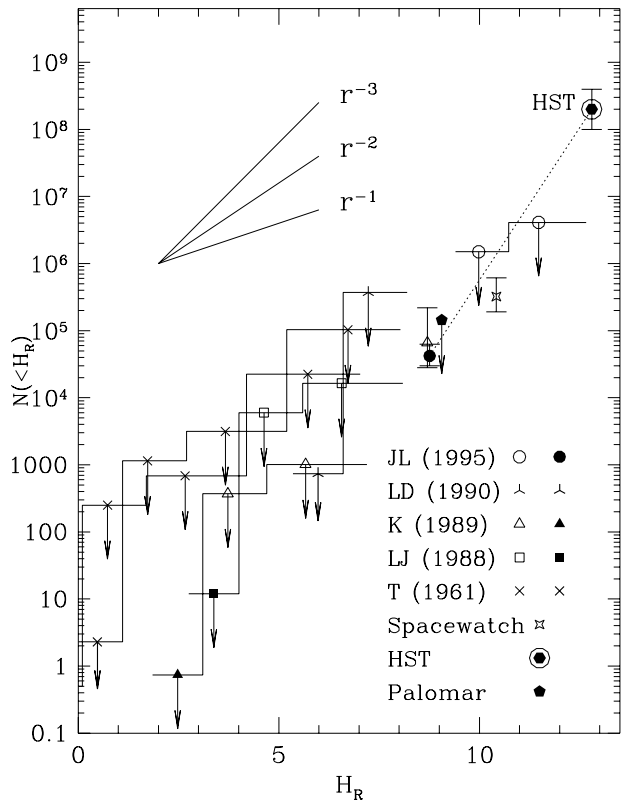,height = 10.5cm}
      \caption{ Constraints on the Kuiper Belt luminosity function,
 		adapted from Irwin et al. (1995).  
 	Filled symbols are direct observational constraints for objects 
        in the Kuiper Belt;
	hollow symbols show model-dependent constraints related to the
  	number of Centaurs interior to Neptune.
	Power-law indices indicate the slopes of various cumulative size 
        distributions in the top center.
	The Palomar point indicates the result of the survey; the target
    	magnitude was in fact $H_R=10$.
              }
         \label{fig:lfunc}
   \end{figure}

Connecting the two positive searches of the Kuiper Belt (the JL 
and HST data points) yields a single power law slope which is
reasonably consistent with all the data (including the upper limits
on the left of the figure).
If true, then there should be $\sim 5$ objects per $10\arcmin \times
10\arcmin$ field at $R\simeq 26$ (the proposed depth of the Palomar
observations).
This pencil-beam survey should thus be able to determine the reality
of this proposed power law, and a negative result would mean that
either the HST result was spurious or that the size distribution
must steepen rather precipitously after $H_R=10$, and cannot be fit 
by a single power law in the range $8 < H_R < 13$.

The orbital properties of the known KBOs are much less well constrained
than may be generally realized.
At the time of writing, of the 39 KBOs ever given provisional
designations, 11 should be considered lost, 18 have been
observed solely at one opposition, and only 11 have multi-opposition
orbits of somewhat good quality (Marsden, 1996, private 
communication).
Therefore, frequent follow-up observations of these objects are
crucial to prevent them from being lost.
Without good orbit determinations, dynamical studies are handicapped
(see Duncan et al. 1995, Morbidelli et al. 1995).
Thus, a second goal of this observation program was to recover as
many known objects as possible to determine astrometric positions,
which were then communicated to the Minor Planet Center.

\vspace*{-0.1in}
%

\section{Observational procedures}

A 2048$\times$2048 thinned Tektronix CCD was used at prime focus
of the 5-m Hale telescope.
The chip has high quantum efficiency (85 -- 90\%  from 550 -- 750
nm) and a fast readout (40 secs with binned pixels of 
0.56\arcsec).
The square field of view is 9.7\arcmin on a side.
Due to the bright sky at Palomar, the Gunn {\it r} filter was used
for the majority of the observations (Thuan and Gunn 1976);
this filter, designed to screen out specific night sky lines, is 
centered at 655 nm with a full-width of 90 nm.

Since KBOs at opposition have retrograde motions of 3--5\arcsec / hour,
integration times were limited to 300 sec to prevent trailing
losses with 0.56\arcsec \ pixels.
For 2\arcsec \ seeing, 300 sec exposures produced a SNR of 4 for objects
with $R\simeq 23$, which is sufficient to recover most known KBOs,
but is clearly insufficient to dectect the faint objects we are
searching for in the pencil-beam survey.
For these objects, we needed to recombine a series of images in
software, assuming a direction and rate for the retrograde
motion; i.e., the same method used in the HST data reduction of
Cochran et al.~(1995).
Fortunately, since our searches were done looking towards
opposition, the predictable retrograde motion always dominates the 
sky rate, and recombinations at different orbital inclinations
(Cochran et al.~1995) were found to be unnecessary.
Therefore, each angular rate corresponds to first order
to different heliocentric distances.
By recombining the frames at a variety of rates, we can search for
objects from 10--60 AU.
Since there appear to be fewer Centaurs per sq.~degree than KBOs
(fainter than $H_R$=8), the most likely discovery is of new KBOs 
in the range 30--40 AU.
Our 300 sec integration time was not optimized to detect Centaurs
in any case, since trailing losses would occur inside Uranus (the poor 
seeing conditions obtained meant that trailing losses between 20 
and 30 AU were not significant).

%
   \begin{table*}
      \caption{Summary of KBO follow-up observations, for all objects for
which images were obtained at 2 or more times during the run.
A number in () following an object indicates that the observation
was the first recovery for that opposition.  (C) indicates that 1993 HA2
is a Centaur, and (P) indicates the 1994 EV3 observation was not
measured due to its poor quality and because it was simultaneously
recovered at Mauna Kea. (N) indicates not found, despite a more than sufficient
limiting magnitude.}
      \label{table}
\begin{center}
\begin{tabular}{lccl}
Date & \# nights & \# open   & \ KBO fields obtained, and results \\ \hline
Mar 1994  &  2   & 0         & \ none \\
Dec 1994  &  3   & 2 hours   & \ 1993 SC (with CFHT help) \\
Feb 1995  &  2   & 6 fields  & \ 1993 FW (3), 1994 EV3(P), 1994 GV9, 1994 JQ1,  1994 JV(N) \\
Jun 1995  &  3   & 2         & \ 1993 HA2 (C), 1993 RO (3), 1994 JS, 1995 KJ1 (1), 1995 KK1(N), 1994 TB (2) \\
Jan 1996  &  4   & 2.5       & \ 1994 GV9(3), 1994 VK8, 1995 DA2, 1995 HM5(2), 1995 YY3(1)  \\
         &       &           & \ 1995 GA7(N), 1995 GJ(N), 1994 JV(N) \\
Apr 1996  &  1   & 3 fields  & \ 1994 GV9, 1995 DC2, 1995 HM5 \\ \hline
\end{tabular}
\end{center}
   \end{table*}

All images involved in the deep search were de-biased and flat-fielded
as usual.
Only the worst cosmic rays were removed, since automated routines
might remove the objects we are looking for.
The data analysis software consists of an 
IRAF\footnote{The Image Reduction and Analysis Facility is
distributed by the National Optical Astronomy Observatories, 
operated by the Association of Universities for Research in Astronomy, Inc.
(AURA) under cooperative agreement with the NSF.}
script which, given
an angular rate and direction, recombines the images by shifting
their pixels and then co-adding them.
The offsets are calculated from the time delay between the start
of any given exposure and one chosen reference frame (usually the 
first).
Thus, all stationary objects will elongate, trailing in the 
direction of recombination; only objects moving at the specified
angular rate will have their signal constructively add into a
single seeing disk.
Experimentation with median filtering the frames before recombination
in order to remove all stationary objects met with mixed success due
to the problem of variable seeing (over the 4--6 hour integration)
causing different point-spread functions.
Since the deep-search fields were selected to have very few background
stars, this refinement produced negligible improvement.

For object recovery, multiple images of the fields were simply blinked
in software to detect the known KBOs.
Astrometric positions were then calculated by computing a full plate
solution using either the HST Guide star catalog, or the APM sky
survey.
Computed positions were generally accurate to sub-arcsecond, neglecting
possible systematic errors in the catalogs.
The positions were reported to the Minor Planet Center for orbit
improvement.
%

\section{Results}

We first summarize the results of the recovery attempts (Table 1).
This portion of the program was quite successful, as evidenced by
the 18 follow-ups, even though the seeing conditions on most of
the nights were poor (1.8\arcsec \ at best).
The February 1995 and April 1996 observations were conducted
as service observing by Tyler Nordgren (Cornell University), and consisted 
of images of KBO fields acquired in the normal course of obtaining
sky flats (a very productive use of this otherwise `dead time').
The KBO 1993 SC was recovered only in conjunction with observations
conducted in early Jan.~1995 at the Canada France Hawaii Telescope
(CFHT) by one of us (JJK).
Considerable time was spent in Jan.~1996 attempting to recover 3
of the KBOs, but despite searches along long arcs to depths one magnitude
fainter than the discovery brightness, 1995 GA7, 1995 GJ, and 1994 JV
could not be found and should be considered lost (and are counted among
the 11 so designated at the end of Sec.~1).

%
   \begin{figure}[htbp]
\epsfig{file=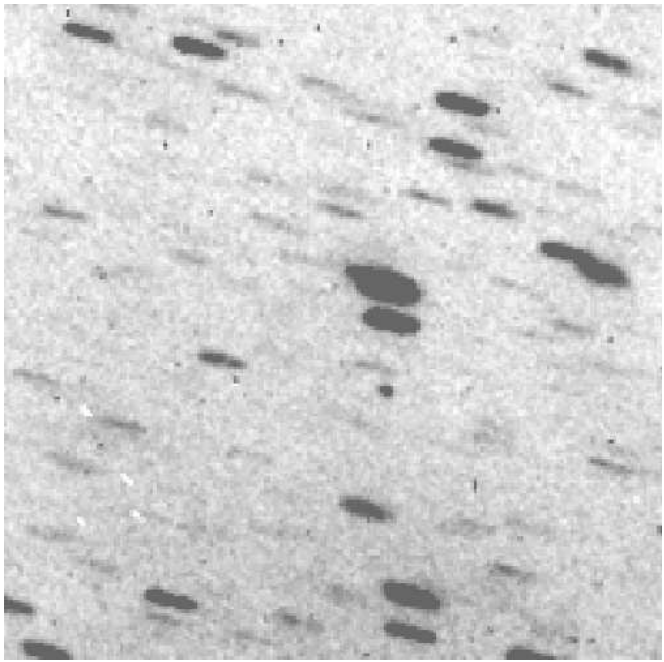,height = 8.7cm}
      \caption{ An $\approx$2.5\arcmin \ portion of the Jan.~1996 deep field 
                showing KBO 
                1995 DA2 (the point-like image near the center of the
                frame, below the two darkest streaks).
		Trails are images of stationary objects, which
  		are smeared due to the recombination algorithm in this
     		shifted and co-added image.
		The many obvious cosmic-rays can be easily
		rejected by their image profile.
              }
         \label{fig:da2}
   \end{figure}

The deep search portion of the program was severly hampered by
consistently poor seeing on those nights for which the dome could
be opened.
Seeing ranging from 1.9\arcsec -- 2.5\arcsec \ was typical, thus
decreasing the limiting magnitude of the deep search.
Only 1 night was usable from the June 1995 run, and 2 nights (of
a single field) were available from the Jan.~1996 observing
run.
Therefore, 0.05 square degrees of sky were searched in total.
The limiting magnitude for SNR=4 objects in the June 1995 field
was $R\simeq 24.8\pm0.2$ (the error being due to the night not
being photometric).
Although many faint main-belt asteroids were detected 
by the detection software, no objects with retrograde
motions slower than 10\arcsec /hour were found.

To insure that the recombination software functioned correctly,
it was decided to conduct the January 1996 search in the field
containing 1995 DA2, so that at least 1 KBO would
exist in the field to be detected.
The two available nights in this run both suffered from seeing
worse than 1.8\arcsec \ and were not photometric.
Nevertheless, we had no trouble in recovering 1995 DA2
($R\approx 23.2$), which was at a SNR $\simeq$~4 on each 300 sec 
exposure and thus easily visible in blinked images.
Fig.~2 shows that the detection algorithm also has no difficulty
(as it should) in finding this object.
The trails on this image are all stationary stars and galaxies,
the point-like object is the KBO, which is the combined result
of 36 images that have been shifted and added at -3.5 \arcsec/hour
(the object's retrograde motion) in the ecliptic.
1995 DA2 has a SNR of 25 in the combined image, and any object
moving at a similar angular rate with a magnitude $R\simeq 25.2\pm0.2$
should come up to a SNR of 4 in the combined image.
The target magnitude of $R\simeq 26$ was not reached due to the
poor seeing and the fact that this caused the deep field to be abandoned 
earlier in the night than was originally planned.

The frames were combined at angular rates between 2 and 
11\arcsec /hour; experiments with artificially implanted objects
showed that an accuracy of 0.5\arcsec /hour was sufficient to
detect the objects, and then the actual shift rate could be
fine-tuned to precisely match the object's motion.
These angular rates correspond to retrograde opposition motions for 
direct circular orbits in the ecliptic plane from 10 -- 60 AU.
A detected object on the first night of the June run could be
recovered from the data from the second night, thus confirming its
motion and allowing an orbit to be computed.
No other (non-asteroidal) moving objects were found on the June fields 
however.

We thus are left with only an upper limit for the population of
KBOs at this magnitude level.
We choose to express this limit as there being $<$1 object
per 0.05 square degrees brighter than $R=25.0 \pm 0.3$, or, using
the assumption of Irwin {\it et al.}, $N<1.4\times 10^5$ KBOs brighter
than $H_R$=9.0, which is plotted on Fig.~1.
The uncertainty in the upper limit is meant to reflect the facts that (1)
none of the nights were photometric, (2) the two nights had different
magnitude limits, and (3) objects weaker than SNR=4 have been recovered
by us, and so this limit might be unnecessarily pessimistic.
Poisson statistics imply $<$20, $<$60, and $<$120 objects per square
degree at the 1, 2, and 3 $\sigma$ levels, respectively.

The single limit thus produced can be seen to be entirely consistent
with all other surveys.
Unfortunately, our target of $H_R$=10 was not obtained due to poor seeing
conditions. 
As can be seen from Fig.~1, if the same upper limit could be placed
at $H_R$=10, then stiff constraints could be placed upon the size 
distribution.
The negative result of this independent survey will help to constrain 
models of the luminosity function of the belt (Weissman and 
Levison 1996).
A single night of good seeing would allow the magnitude limit to be improved
dramatically.

\vspace*{-0.1in}

%

\section{Conclusions}

We have attempted to address two major issues in
Kuiper Belt research: (1) improving the orbits of known KBOs
for dynamical studies, and (2) searching for fainter
(and thus smaller) objects in order to improve knowledge of the
size distribution.
The objects recovered work have had their orbits constrained 
considerably, particularly for 1994 TB (Minor Planet Electronic
Circular 1995-M07) and 1995 HM$_5$ (MPEC 1996-C05). 
The observations/orbits for first opposition objects
can be found in Minor Planet Circulars 25494/25514 (1995 KJ$_1$) and
26660/26724 (1995 YY$_3$). 
An improved orbit for 1995 HM$_5$ from the April 1996 
observations appears in MPC 27122. 
The upper limit on faint objects from the deep survey in consistent
with previous surveys.

\begin{acknowledgements}
We thank Joseph A. Burns, Anita Cochran, David Hanes, 
Martha Haynes, Terry Herter, Hal Levison, Brian Marsden, Tyler Nordgren, 
Marco Scodeggio, and Scott Tremaine for their assistance in various 
phases of this work.
This work was supported by NASA Grant NAGW-310.
\end{acknowledgements}


\end{document}